\documentclass[a4paper,11pt]{article}
\usepackage{jheppub} 
\usepackage{lineno}
\usepackage{hyperref}
\usepackage{multirow}

\title{\boldmath Exploring Low Energy Excess in MINER with sapphire detectors using Convolutional Variational Autoencoder (CVAE)}

\author[1,2]{D.~Mondal}
\affiliation[1]{National Institute of Science Education and Research, Jatni-752050, Odisha, India}
\affiliation[2]{Homi Bhabha National Institute, Training School Complex, Anushakti Nagar, Mumbai, 400094, Maharashtra, India}
\author[3]{W. Baker}
\affiliation[3]{Department of Physics and Astronomy, Texas A\&M University, 578 University Dr, College Station, 77840, TX, US}
\author[1,2]{M. Chaudhuri}
\author[4]{J. B. Dent}
\affiliation[4]{Department of Physics, Sam Houston State University, 1905 University Ave, Huntsville, TX, 77340, US}
\author[3]{B. Dutta}
\author[5]{V. Iyer}
\affiliation[5]{SNOLAB, Creighton Mine No. 9, 1039 Regional Road 24, Sudbury, ON P3Y 1N2, Canada}
\author[6]{A. Jastram}
\affiliation[6]{Department of Mechanical Engineering, Texas A\&M University, 202 Spence St, College Station, 77840, TX, US}
\author[1,2]{\ \\V. K. S. Kashyap}
\author[5]{A. Kubik}
\author[7]{K. Lang}
\affiliation[7]{Department of Physics, The University of Texas at Austin, 2515 Speedway, Austin, TX, 78712, US}
\author[3]{R. Mahapatra}
\author[3]{S. Maludze}
\author[3]{N. Mirabolfathi}
\author[3]{M. Mirzakhani}
\author[1,2]{B. Mohanty}
\author[8]{H. Neog}
\affiliation[8]{School of Physics \& Astronomy, University of Minnesota, Minneapolis, MN, 55455, US}
\author[9]{J.~L.~Newstead}
\affiliation[9]{ARC Centre of Excellence for Dark Matter Particle Physics, School of Physics, The University of Melbourne, Parkville VIC 3010, Australia}
\author[3]{M. Platt}
\author[3]{S. Sahoo}
\author[10]{\ \\J. Sander}
\affiliation[10]{Department of Physics, University of South Dakota, 414 E Clark St, Vermillion, SD, 57069, US}
\author[3]{L. E. Strigari}
\author[11]{J. Walker}
\affiliation[11]{Department of Physics, Sam Houston State University, 1905 University Ave, Huntsville, TX, 77340, US}

\collaboration{MINER Collaboration}

\emailAdd{bedanga@niser.ac.in}
\emailAdd{dipanwita.mondal@niser.ac.in}
\abstract{
As cryogenic detectors push toward ever-lower energy thresholds, their sensitivity is increasingly constrained by a persistent low-energy background known as the low-energy excess (LEE). 
We report observation of LEE in the MINER experiment using a sapphire ($\mathrm{Al_2O_3}$) detector at energies around 200 eV, with the excess reproducibly reappearing after each non-operational warm-up period. To address this limiting background, we implement an unsupervised convolutional variational autoencoder (CVAE) framework that identifies anomalous events through a reconstruction-based anomaly score.

For the first time in a pulse-shape driven analysis, we uncover a significant deviation in the rise-time of LEE events relative to Monte Carlo simulated ideal signals. Using this feature, we develop a discrimination pipeline based on rise-time selection. This method achieves up to 53\% rejection of LEE events, corresponding to an expected sensitivity improvement of nearly 10\% for MINER at HFIR.

These findings are consistent with a scenario in which a substantial fraction of the LEE originates from bulk-related defects or microfractures within the detector crystal, while leaving room for additional detector-related contributions. Our result provides a powerful, data-driven pathway for mitigating LEE and enhancing the discovery potential of next-generation cryogenic experiments.}


\begin{document}
\maketitle
\flushbottom

\section{Introduction\label{sec1:intro}}
Low-threshold cryogenic detectors are well-suited for the detection of very small recoil energy, in the range of tens of eV to a few keV that could be produced by rare events like dark matter (DM) or the coherent elastic neutrino-nucleus scattering (CE$\nu$NS) processes. Recent progress in the cryogenic experiments has given us access to the sub-eV energy region, unlocking a new background in the low-energy range, known as low-energy excess (LEE). Experiments such as EDELWEISS \cite{LEE_EDELWEISS_2019} and CRESST \cite{LEE_CRESST_2019} observed it first in 2019 in Ge and $\mathrm{CaWO_4}$ detectors, respectively. In recent years, several studies have investigated this phenomenon experimentally, including EDELWEISS \cite{EDELWEISS_2020,EDELWEISS_2023}, CRESST \cite{LEE_CRESST_confpros}, NUCLEUS \cite{LEE_nucleus_2026}, Ricochet \cite{Ricochet_2023}, SuperCDMS \cite{LEE_SuperCDMS}, and TESSERACT \cite{LEE_TESSERACT_bulk,TESSERACT_2025,TESSERACT_2026}, which utilize phonon readout techniques, as well as CONNIE \cite{CONNIE_2024}, DAMIC \cite{DAMIC_2021,DAMIC_2023}, and SENSEI \cite{SENSEI_2025}, which employ charge readout methods. The striking feature of this background, which has been observed in both above and underground experiments, is that near the operation threshold, the spectra rise very sharply. Moreover, dedicated studies show the rate decreases with time after initial cooling and reappears again after heating and recooling \cite{CRESST_LEE_2024, LEE_CRESST_confpros,TESSERACT_2026, LEE_nucleus_2026}. Furthermore, the shape of these spectra varies between all the experiments and between different detector materials, which rejects the possible origin of this excess from DM particles \cite {LEE_CRESST_2019}. A detailed summary of these observations across all the experiments can be found in \cite{excess_workshop} and \cite{LEE_annual_review}. A discrepancy between the simulated background and the data has been also observed at Mitchell Institute Neutrino Experiment at Reactor (MINER) using a sapphire detector \cite{MINER_CEvNS}, suggesting a possible contribution from LEE events (discussed in detail in Sec. \ref{sec2:lee_miner}).

 Several hypotheses are proposed, and some of them are experimentally tested as possible mechanisms behind the generation of LEE, including the relaxation of stress in the detector holdings \cite{nature_2024} or, in the aluminum fins \cite{LEE_al_relaxation}, relaxation of defects produced inside the crystals due to radiation or fast neutrons induced from cosmogenic backgrounds \cite{microfracture_LEE,radiation_LEE,TESSERACT_2026}. Though any conclusive results are yet to be found, the TESSERACT collaboration in their recent study has indicated that the effect could be a detector-related bulk effect that scales with volume \cite{LEE_TESSERACT_bulk}. 

 The expected signal rates for experiments such as DM searches and CE$\nu$NS peak at low energies, precisely where the LEE dominates and limits detector sensitivity. Consequently, understanding and mitigating this background remains a critical challenge for fully exploiting the potential of low-threshold experiments.

The application of machine learning techniques to physics problems has grown significantly in recent years. Convolutional neural networks (CNNs), a class of supervised learning models, have been widely used for tasks such as classification and regression, as demonstrated in Refs.~\cite{CNN_XENON_2019,CNN_NEXT,CNN_CVAE_DARWIN,CNN_DM_halo_2019}. In addition, unsupervised learning methods have been extensively employed in event simulation and anomaly detection at the LHC \cite{CVAE_LHC2_2021,CVAE_LHC1_2021}. In this context, an unsupervised convolutional variational autoencoder (CVAE)-based approach provides a powerful framework to capture shape-based deviations, using measures such as anomalousness or reconstruction loss. In this study, we train a CVAE network on Monte Carlo (MC) data and apply it to data dominated by LEE events, where an increase in anomaly is expected. Although supervised algorithms are generally more suitable for classification tasks, the pulse shape of LEE events is not yet well understood; therefore, an unsupervised approach is more robust and helps avoid potential biases.

In this article, section~\ref{sec2:lee_miner} describes the observed LEE at MINER in sapphire detector.  A description of CVAE and training procedure, as well as preparation of the MC is given in section~\ref{sec3:CVAE_network}. Section~\ref{sec4:results_analysis} contains the analysis procedure and results. Finally, conclusions are drawn in section~\ref{sec5:conclusion}.

\section{Time dependence LEE rate at MINER\label{sec2:lee_miner}}
MINER, a reactor-based experiment at Texas A\&M University, utilizes a 1 MW$_{\mathrm{th}}$ low-enriched (20\% $^{235}\mathrm{U}$) reactor located approximately 4 m from the detector to detect CE$\nu$NS. In a recent study, MINER reported an upper limit on CE$\nu$NS sensitivity using a 72 g sapphire detector system \cite{MINER_CEvNS}. In the present work, we analyze the same dataset used in that study. A detailed description of the experimental configuration, including the payload and readout system, is available in \cite{MINER_CEvNS}.

\begin{figure}
    \centering
    \includegraphics[width=0.9\linewidth]{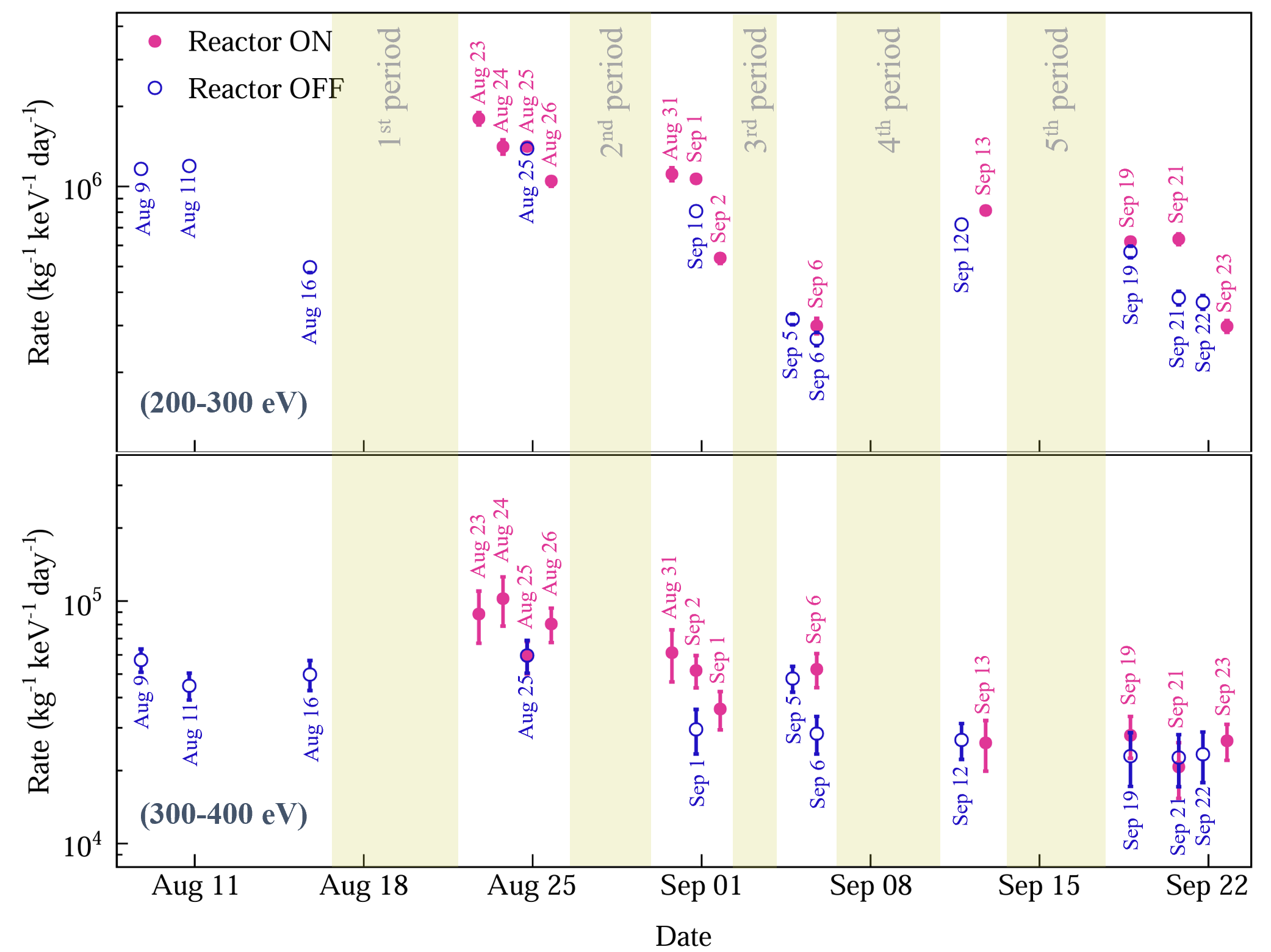}
    \caption{  \label{fig:LEE_reactor_on_off}Single-scatter spectrum in the low-energy region, collected over multiple data-taking periods, demonstrating a systematic increase then decrease in rate in the first two energy bins (1$^{\mathrm{st}}$ bin: 200–300 eV and 2$^{\mathrm{nd}}$ bin: 300–400 eV) with increasing time after each non-operational warm-up period for reactor-on (red markers) and reactor-off (blue markers) conditions. The yellow shaded region denotes the non-operational period during the data-taking campaign.}
\end{figure}

In reactor-based CE$\nu$NS experiments, such as CONUS$+$~\cite{COUNSplus}, TEXONO~\cite{TEXONOresult}, CONNIE~\cite{CONNIE}, NUCLEUS~\cite{NUCLEUS_2019,NUCLEUS_2025}, Ricochet~\cite{Ricochet_comissioning}, and $\nu$GEN~\cite{NUGEN}, the reactor-OFF spectrum is used to characterize environmental and cosmogenic backgrounds, while the reactor-ON spectrum includes both the OFF background and reactor-correlated contributions. Therefore, the ON–OFF difference is expected to isolate the reactor-correlated component, provided that the signal expectation is very low. However, a mismatch is observed at MINER between the Geant4-simulated reactor background spectrum and the data near the threshold ($\sim$200 eV), even as the expected CE$\nu$NS signal rate was very low  ($\sim$0.14 events/kg/day). In contrast, good agreement is found above 400 eV. This discrepancy motivates the investigation of a possible contribution from LEE events. For this study, we select single-scatter (signal-like) events in the 200–400 eV energy range, as described in Ref.~\cite{MINER_CEvNS}, for reactor-ON and OFF conditions. We further subdivide this mismatched region into two intervals: 200–300 eV and 300–400 eV, and examine the time evolution of the event rates (shown in Fig.~\ref{fig:LEE_reactor_on_off}).

The data-taking campaign began in early August 2022. As the reactor facility was not available continuously, reactor-ON and OFF (background) data were collected in several sub-datasets. Starting from August 9, the dataset used in Ref.~\cite{MINER_CEvNS} was extended, and the campaign continued until September 23, 2022. During this period, there were five non-operational intervals, during which the refrigerator temperature increased to approximately 4 K, compared to the base temperature of 8 mK. Figure~\ref{fig:LEE_reactor_on_off} shows that immediately after the first warming period, a significant increase in both reactor-ON and reactor-OFF event rates is observed, particularly in the 200–300 eV range, while the 300–400 eV interval also exhibits a noticeable rise. Following this, during subsequent non-operational periods, the 200–300 eV interval shows a considerable increase in rate that gradually decays over time. In contrast, the 300–400 eV interval exhibits an overall decrease, eventually stabilizing toward the end of the run.

By contrast to the background data collection campaign, the reactor-ON data were collected only after the first warm-up period. Since most of the reactor-ON data were acquired shortly after the warm-up periods, a larger contribution from LEE events is expected in the reactor-ON dataset compared to the reactor-OFF data. As a consequence, a mismatch between the simulated spectra and the reactor-ON minus reactor-OFF spectra is observed, as shown in~\cite{MINER_CEvNS} (Fig.~10). LEE is generally believed to be detector-related and therefore would be partially suppressed by the ON–OFF subtraction. However, its time-dependent behavior, combined with the fact that reactor-ON and reactor-OFF data were recorded during different periods, implies that a residual LEE contribution may persist even after subtraction.

\section{Study of LEE using unsupervised learning\label{sec3:CVAE_network}}
 The presence of LEE in the reactor-ON minus OFF spectrum can limit the sensitivity of reactor-based CE$\nu$NS experiments, particularly near the energy threshold. Therefore, in addition to understanding the origin of LEE, developing effective mitigation strategies is of utmost importance. Motivated by this goal, we attempt to discriminate LEE events using an unsupervised CVAE, based on reconstruction loss relative to ideal pulse shapes, since the characteristic shape-based features of LEE events are not yet well understood. 
\subsection{Convolutional Variational Autoencoder (CVAE)\label{sec3.1:CVAE}}
Autoencoders (AEs) \cite{autoencoders_2021} are a type of unsupervised neural network that attempt to reconstruct the output as close to the input as possible. An AE mainly consists of three parts: an encoder, a latent space and a decoder. An encoder compresses the input information of a data vector ($\mathrm{\boldsymbol{x}_{in}\in \mathbb{R}^m}$) into a lower-dimensional latent space $\mathrm{\boldsymbol{z}\in \mathbb{R}^p}$ ($m\gg p$ ). On the other hand, the decoder produces an output, $\mathrm{\boldsymbol{x}_{out}=}g\mathrm{(\boldsymbol{z})}$, by reconstructing the input from the compressed latent space. This is done by optimizing a reconstruction loss function, commonly defined for a sample of $N$ inputs as
\begin{equation}
    L_\mathrm{reco}=\frac{1}{N}\sum_i^N(x_{i,\mathrm{in}}-x_{i,\mathrm{out}})^2.
    \label{eq:recon_loss}
\end{equation}
The lower the loss, the better the reconstruction. An AE is trained on a data set with ideal event pulses; if applied on a data set with both ideal and LEE-dominated events, it is expected to have a low reconstruction loss for ideal events and a high loss if any shape-based anomaly exists. In a non-ideal case, the encoder would place the event in an improbable location in the latent space from which the decoder cannot reconstruct it properly \cite{AE_anomaly_detection}, resulting in a high loss score. 

Variational Autoencoders (VAEs) differ from traditional autoencoders (AE) \cite{VAE_2022}. Rather than mapping an input to a single point in the latent space, a VAE represents the input as a Gaussian distribution defined by a mean $\mu$ and variance $\sigma^2$, i.e., $\mathcal{N}(\mu,\sigma^2)$. Each dimension of this latent space is intended to capture a specific feature of the data.
During decoding, samples are drawn randomly from these distributions to reconstruct the input. This probabilistic approach results in a more structured and continuous latent space, enabling smoother sampling and more realistic data generation.

In comparison to the reconstruction loss defined in Eq.~\ref{eq:recon_loss}, training a VAE involves augmenting the loss function with a Kullback–Leibler (KL) divergence term, leading to the following formulation:

\begin{equation}
    L_\mathrm{loss}=L_\mathrm{reco}+\beta D_{KL}
    \label{eq:loss_tot}
\end{equation}
where for a $q$-dimensional latent space with $j^\mathrm{th}$ normal distribution having mean and variance $\mu_j$ and $\sigma_j$ respectively, the KL divergence term is defined as
\begin{equation}
    D_{KL}=\frac{1}{2}\sum_{j=0}^q [ \sigma^2_j+\mu_j^2-\log(\sigma_j^2)-1]
    \label{eq:divergence_loss}
\end{equation}
This divergence term serves as a regularization mechanism, ensuring that the encoder does not map inputs to arbitrary or highly disjoint regions of the latent space. In the absence of this constraint, the model effectively reduces to a standard autoencoder, which generally performs poorly at generating new data \cite{KL_divergence}. The constant factor $\beta$ controls the influence of the regularisation term. The final loss score will be a trade-off between the reconstruction loss and the regularisation. Although a higher value of $\beta$ will lead to a less detailed reconstruction of the original pulse, a lower value of $\beta$ would lose any control over the latent space, making VAE a normal AE.

A CVAE is a variant of the VAE architecture that incorporates convolutional layers to enhance feature extraction while effectively reducing input dimensionality \cite{CVAE}.

\subsection{Data preparation\label{sec3.2:data_prep}}
In this study, two sets of data are used: one is real data of single-scatter (signal-like) events, and the other is MC-generated ideal signal events (10,000 events). The CVAE model is trained on MC data and is applied to real data. 

Single-scatter events are defined as those depositing energy exclusively in the primary 4 mm detector, whereas events producing signals in multiple detectors are classified as background. A detailed description of the event selection pipeline is provided in~\cite{MINER_CEvNS}. Since single-scatter events constitute the observable of interest, it is essential to quantify the fraction of LEE events that pass the selection criteria and potentially degrade the sensitivity in the low-energy region. As discussed in Sec.~\ref{sec2:lee_miner}, a deviation between data and simulation is observed in the 200–400 eV energy range. Therefore, a background data set that shows a high LEE rate just after the warm-up period, consisting of real single-scatter events in the 200–500 eV range, is constructed for further analysis.

\begin{figure}[htbp]
    \centering
    \begin{minipage}{0.34\columnwidth}
        \centering
        \includegraphics[width=\linewidth]{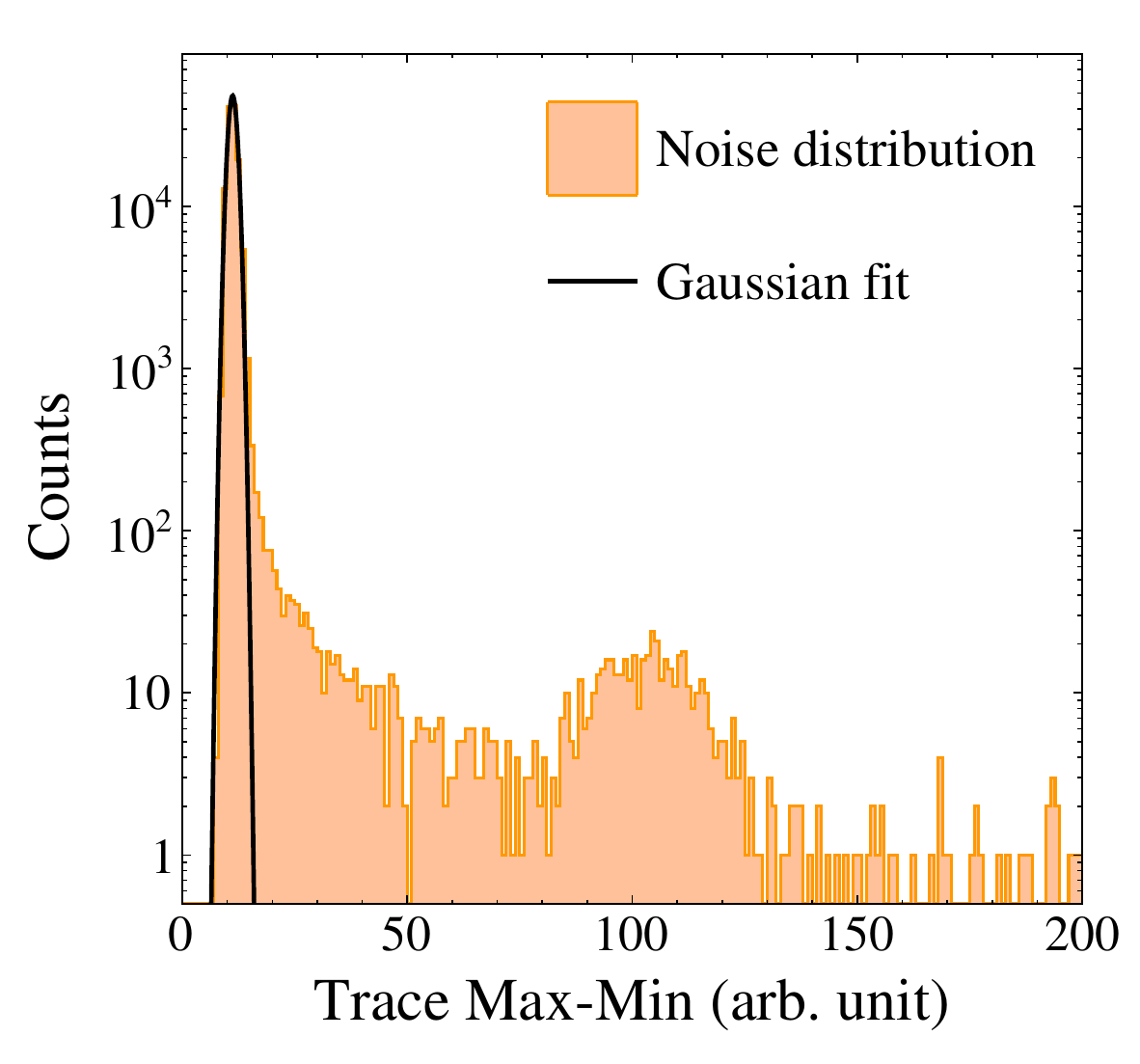}
        \label{fig:noise_sel}
    \end{minipage}
    \begin{minipage}{0.652\columnwidth}
        \centering
        \includegraphics[width=\linewidth]{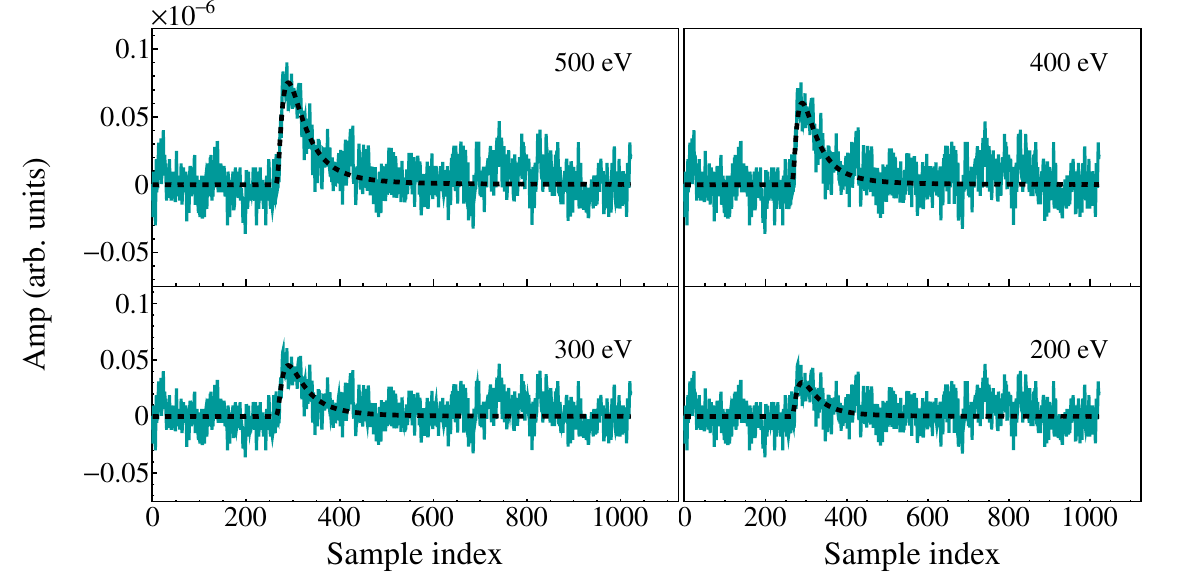}
        \label{fig:mc_example}
    \end{minipage}
    \caption{\label{fig:noise_sel_MC_exmp}\textit{Left}: Selection of good-quality noise traces based on the trace (Max.–Min.) distribution. The black curve represents the Gaussian fit.
    \textit{Right}: MC simulated pulses (green) in the primary sapphire detector at 200 eV, 300 eV, 400 eV, and 500 eV, compared with the corresponding pulse templates (black dashed line).}
\end{figure}

The MC data set is generated by injecting pulse templates into randomly selected noise traces from the same background data set as real data that satisfy the selection criteria. In general, high-quality noise traces are obtained from randomly triggered segments of continuous data. To minimize the possibility of accidental pulse contamination in these traces, the distribution of the maximum minus minimum (Max.–Min.) values of each trace is examined as shown in the left plot in Fig.~\ref{fig:noise_sel_MC_exmp}. Traces without pulses are expected to populate the lower region of this distribution. A Gaussian fit is performed in this region, and only traces within the range $\mu \pm 3\sigma$ are selected as good noise candidates. To ensure consistency with real data, the same energy range 200–500 eV is selected. This energy range is then divided into 100 bins, each containing 100 events. The right panel of Fig.~\ref{fig:noise_sel_MC_exmp} shows representative examples of simulated pulses with randomly selected good noise traces. In this study, we assume that, in an ideal scenario, true signal events should closely resemble these MC-generated pulses.
\subsection{Network architecture\label{sec3.3:CVAE_archi}}
\begin{figure}
    \centering
    \includegraphics[width=\linewidth]{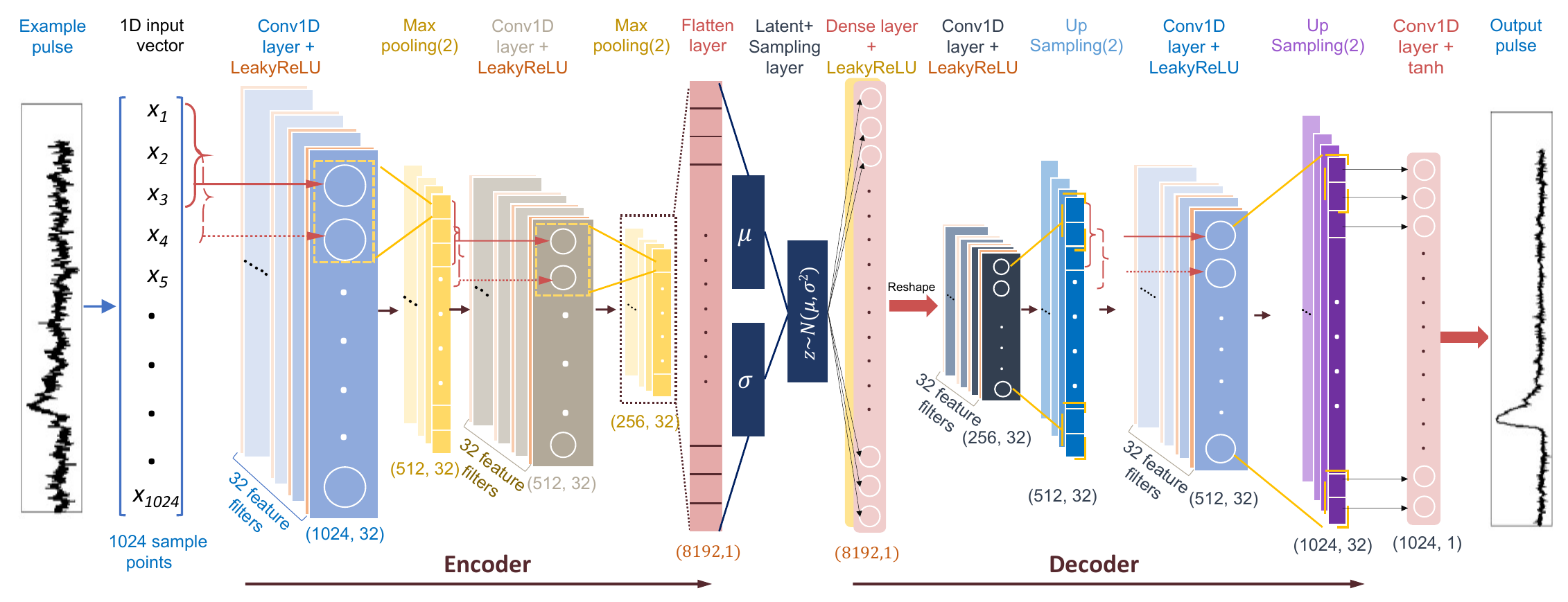}
    
    \caption{\label{fig:cvae_network}
Schematic representation of the CVAE architecture. The encoder compresses the input pulse into a latent representation characterised by mean and variance parameters, each with shape 256, from which a latent vector is sampled using the reparameterization trick. The decoder reconstructs the pulse from this latent vector using convolutional and upsampling layers. The network is trained by minimizing a combination of reconstruction loss and Kullback–Leibler divergence as defined in~\ref{eq:loss_tot}. The displayed input and output pulses correspond to the true input waveforms and their reconstructed counterparts.}
\end{figure}

\begin{table}
\centering
\small
\begin{tabular}{|c|c|c|}
\hline
\textbf{Section} & \textbf{Layer / Operation} & \textbf{Output Shape} \\
\hline

\multirow{5}{*}{Encoder} 
& Input Signal & $(T,1)$ \\
& Conv1D(32) + LeakyReLU & $(T,32)$ \\
& MaxPooling1D(2) & $(T/2,32)$ \\
& Conv1D(32) + LeakyReLU & $(T/2,32)$ \\
& MaxPooling1D(2) & $(T/4,32)$ \\
& Flatten & $(T/4 \times 32)$ \\

\hline

\multirow{3}{*}{Latent} 
& Dense ($z_{\mu}$) & $(256)$ \\
& Dense ($z_{\log\sigma^2}$) & $(256)$ \\
& Sampling ($z$) & $(256)$ \\
\hline

\multirow{6}{*}{Decoder} 
& Dense + LeakyReLU & $(T/4 \times 32)$ \\
& Reshape & $(T/4,32)$ \\
& Conv1D + LeakyReLU & $(T/4,32)$ \\
& UpSampling1D(2) & $(T/2,32)$ \\
& Conv1D + LeakyReLU & $(T/2,32)$ \\
& UpSampling1D(2) & $(T,32)$ \\
& Output Conv1D (tanh) & $(T,1)$ \\
\hline


\hline

\end{tabular}
\caption{\label{tab:cvae_architecture}
Summary of the CVAE architecture used in this analysis. The table lists the sequence of layers in the encoder and decoder, along with their corresponding output shapes. The model compresses the input pulse of length $T=1024$ into a 256-dimensional latent representation and reconstructs it using symmetric convolutional and upsampling blocks. The training loss consists of a reconstruction term (mean squared error) and a Kullback–Leibler divergence regularization term.}
\end{table}

The architecture of the CVAE network used in this analysis is shown in Fig.~\ref{fig:cvae_network} and is implemented using \texttt{TensorFlow}~\cite{tensorflowl_2016} (v2.13.1). A summary of the architecture is also provided in Table~\ref{tab:cvae_architecture}. The CVAE network consists of several subcomponents: encoder and decoder blocks, along with a sampling layer that enables stochastic sampling from the latent space. Both the encoder and decoder blocks comprise one-dimensional convolutional layers followed by activation functions and either pooling or upsampling layers.

Convolutional layers extract features from the input by applying kernels over the input vector. Each convolutional layer employs multiple filters to capture different characteristics of the pulse shape, such as amplitude, rise-time, and decay structure. The activation function used in this study is the Leaky Rectified Linear Unit (Leaky ReLU)~\cite{relu_2013}, defined as
\begin{equation}
f(x) =
\begin{cases}
\alpha x & \text{if } x < 0 \\
x & \text{otherwise}
\end{cases},
\end{equation}
where $\alpha$ is a constant parameter that prevents neurons from becoming inactive when encountering negative inputs. This property helps preserve small but relevant features in the signal.

In the encoder block, max-pooling layers with a pooling size of 2 are applied after convolutional layers to reduce dimensionality and compress features into the latent space by retaining the maximum value within each pooling window. In the decoder block, upsampling layers with a factor of 2 are used to restore the temporal resolution by duplicating values along the time axis, thereby reconstructing the original signal dimensionality.

The network takes as input a pulse of shape $(1024 \times 1)$, where each trace represents the temporal evolution of the signal amplitude. In the encoder, feature extraction is performed using one-dimensional convolutional layers with a kernel size of 3 and 32 filters. The first convolutional layer transforms the input to an output of shape $(1024 \times 32)$, which is subsequently downsampled by a factor of 2 using max-pooling, resulting in $(512 \times 32)$ shape. A second convolutional block further reduces the dimensionality to $(256 \times 32)$. These features are then flattened and mapped to a 256-dimensional latent space through dense layers that parameterize the mean and log-variance, followed by a sampling layer implementing the reparameterization trick.

The decoder mirrors the encoder architecture. Starting from the latent vector, a dense layer expands the representation, which is reshaped into $(256 \times 32)$. Two convolutional blocks, each followed by upsampling layers, progressively restore the temporal resolution, ultimately reconstructing the signal to its original shape of $(1024 \times 1)$.

\subsection{Training\label{sec3.4:training_network}}
\begin{figure}
    \centering
    \includegraphics[width=0.6\linewidth]{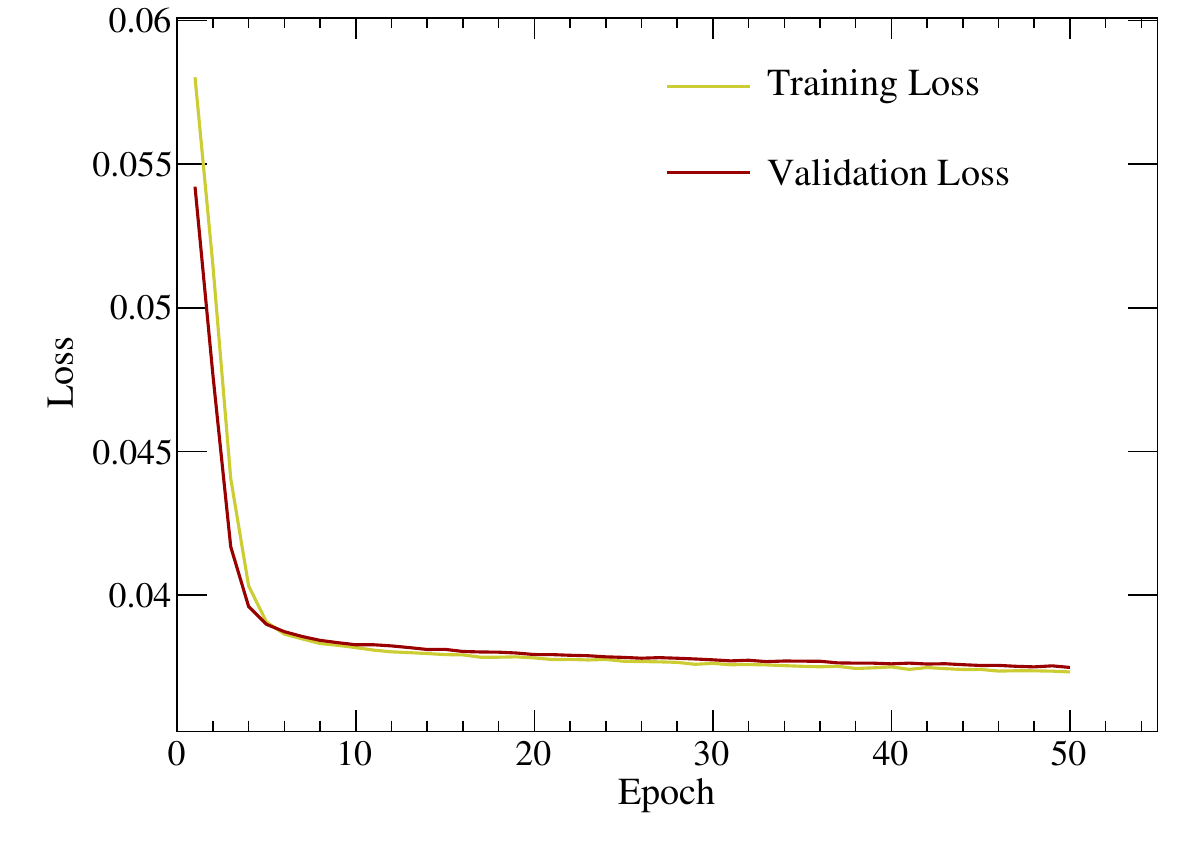}
    \caption{Total loss over 50 epochs, defined in Eq.~\ref{eq:loss_tot}, for the training and validation sets of the CVAE model. Both losses decrease rapidly during the initial epochs and begin to saturate after approximately epoch 20, indicating convergence of the model and successful learning of the relevant data features. The close agreement between the training and validation losses suggests good generalization to unseen data.
    \label{fig:train_val_loss}}
\end{figure}
The CVAE architecture described in Sec.~\ref{sec3.3:CVAE_archi} is trained using 10,000 MC events for 50 epochs with a batch size of 64. The dataset is split into training and validation subsets in an 80\%–20\% ratio, respectively. Optimal performance is achieved when all convolutional layers are followed by Leaky ReLU activation functions with $\alpha = 0.05$, which help preserve small signal features while preventing inactive neurons. The network is trained with a learning rate of $10^{-4}$, and the loss function comprises a reconstruction term and a Kullback–Leibler divergence term, weighted by a regularization factor $\beta = 5 \times 10^{-2}$.

Figure~\ref{fig:train_val_loss} shows the evolution of the training and validation losses, which remain comparable throughout the training process, indicating stable and well-converged learning. Once trained, CVAE assigns higher anomaly scores (reconstruction loss) to events that deviate from the MC distribution, while events resembling ideal (MC-like) pulses yield lower scores.

\section{Analysis and Results \label{sec4:results_analysis}}
\begin{figure}
    \centering
    \includegraphics[width=0.6\linewidth]{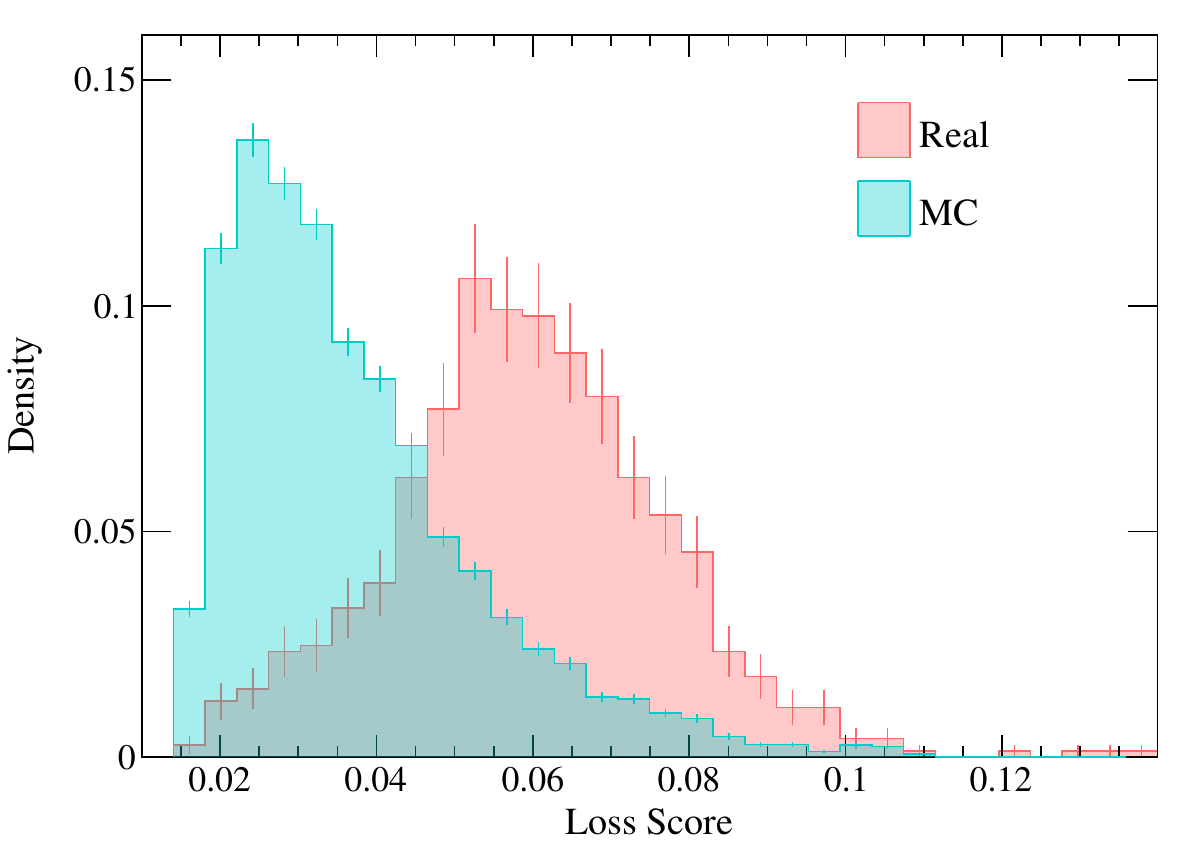}
    \caption{Distribution of loss score as defined in Eq.~\ref{eq:loss_tot} for MC (Cyan) and Real (Red) data.}
    \label{fig:density_loss_real_fake}
\end{figure}
Figure~\ref{fig:density_loss_real_fake} shows the distribution of reconstruction loss scores for MC and real data after being processed through the trained CVAE network. The MC dataset consists of 10,000 events, while the real dataset contains 727 events. For a fair comparison, the distributions are normalized and presented as probability densities. As shown in Fig.~\ref{fig:density_loss_real_fake}, the majority of MC events are concentrated in the low-loss region, indicating good reconstruction of ideal pulse shapes. The tail of the MC distribution is primarily populated by low-energy events, where the signal-to-noise ratio is relatively poor. In contrast, most of the real data events are distributed toward higher loss values, suggesting a deviation from the MC-like pulse shapes.

\subsection{Anomaly detection\label{sec4.1:anomaly_detection}}
\begin{figure}
    \centering
    \includegraphics[width=\linewidth]{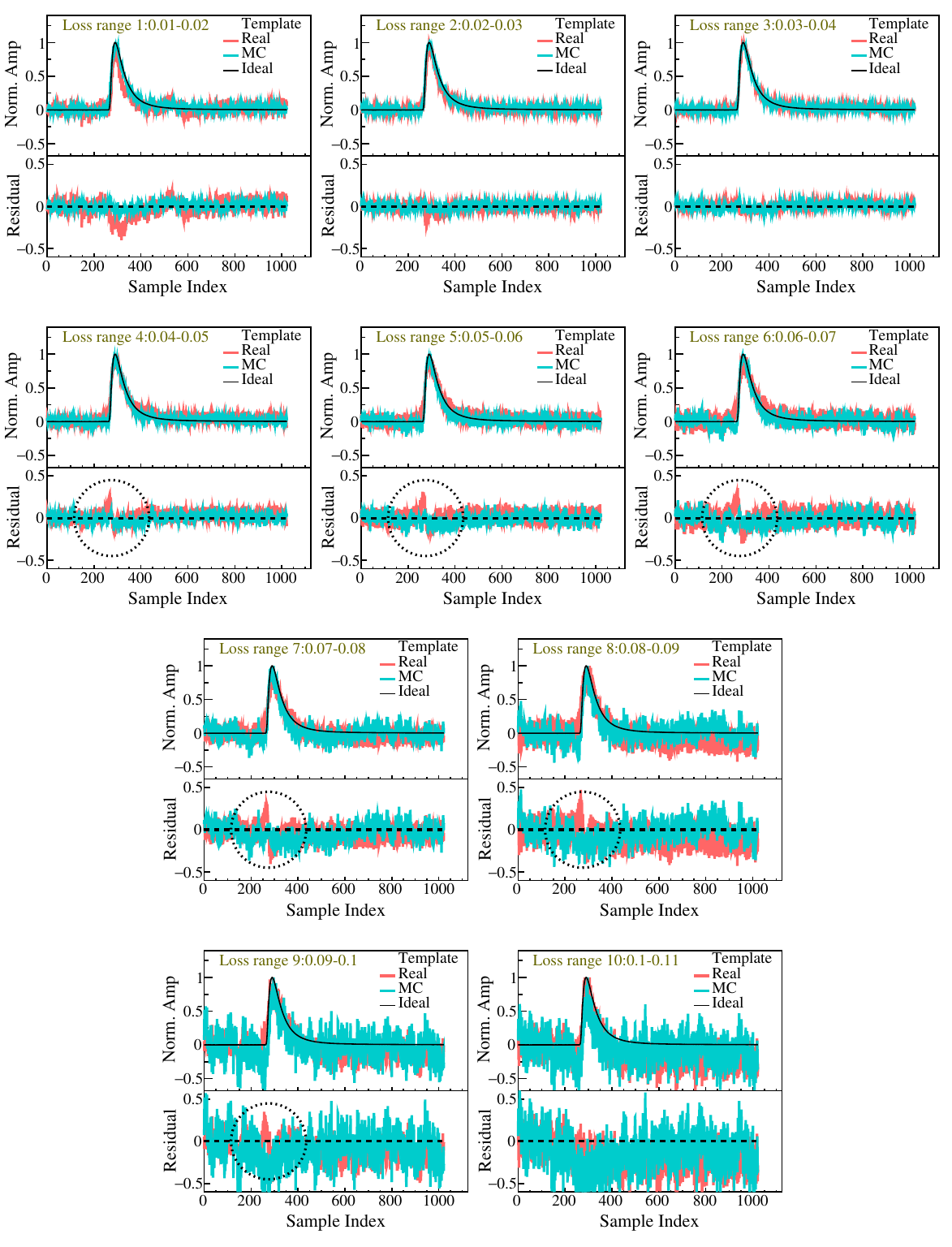}
    \caption{Comparison of templates corresponding to different reconstruction loss score ranges derived from MC and data. The relevant loss score interval for each panel is indicated in green. In every plot, the upper panel compares the real-data (red) and MC (green) templates with the ideal template (black). The lower panel shows the residuals obtained after subtracting the ideal template. The black dotted line represents the baseline for reference, while the black dotted circles highlight the regions in the residual spectra where the real data deviates from the MC prediction.}
    \label{fig:anomaly_temp_real_fake}
\end{figure}

Although most single-scattered events are concentrated in the higher loss-score region rather than the low-loss region, the data and MC distributions are not completely disentangled. To further investigate potential template-level differences, the full loss-score range, starting from 0.01 to 0.11, is divided into ten sub-ranges. For each loss-score interval, all events are selected, averaged, and normalized to construct representative templates. The same procedure is applied to both MC and real data. Ideal templates are also generated using high-energy pulses following an identical method.

In Fig.~\ref{fig:anomaly_temp_real_fake}, the loss score increases consecutively by 0.01 from the upper-left panel of the top row to the lower-right panel of the bottom row. Each panel shows a comparison between MC and real data templates with respect to the ideal template. A noticeable excess-like feature is observed in the residuals of the real data in the sample index range of 200–400, while the MC residuals remain consistent with the baseline. However, this feature appears only in templates corresponding to loss scores between 0.04 and 0.1.

Although a deviation from MC is also visible in the left-most panel of the top row, the statistical uncertainty in this low loss-score region is large due to the small number of real events. Therefore, it is unclear whether this deviation has a physical origin or is merely a statistical fluctuation. A similar argument applies to the last two panels of the bottom row. In contrast, the second and third panels of the top row show good agreement between data and MC, indicating that in the low loss-score regime, the simulation accurately reproduces the real data.

As the loss score increases, a slow component emerges in the pre-pulse region, influencing the rise-time of the templates. This behavior may indicate a potential signature of LEE events.

\subsection{LEE: rise-time estimation\label{sec4.2:rise_time_cal}}
\begin{figure}
    \centering
    \includegraphics[width=0.8\linewidth]{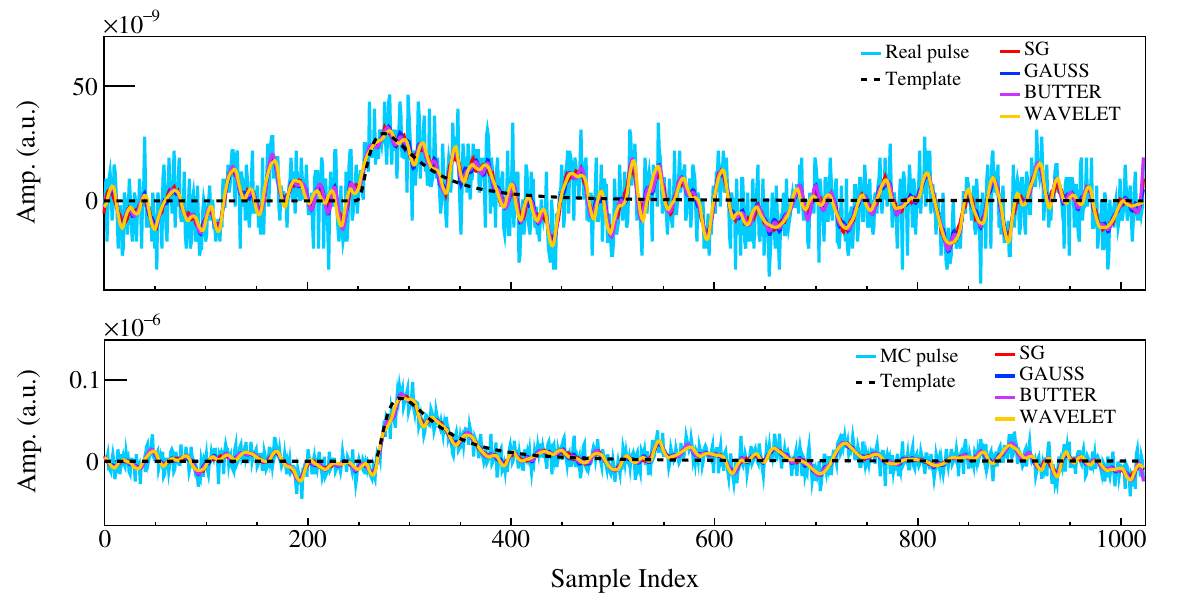}
    \caption{ Example real pulse compared with the ideal template (black) and filtered pulses obtained using Savitzky–Golay (red), Butterworth (violet), Gaussian (yellow), and wavelet (cyan) filters. The upper panel shows a real pulse, while the lower panel shows a corresponding MC pulse.}
    \label{fig:filtered_pulse_fake_real}
\end{figure}
To study and compare the rise-time of real and MC-simulated templates, the pulses are processed through filters that suppress high-frequency noise while preserving the overall waveform shape. Four filtering techniques are applied to compare the results: Gaussian and Savitzky–Golay (SG) filters in the time domain, Butterworth in the frequency domain, and wavelet-based methods in the time–frequency domain.

A very common choice is the Gaussian filter, which is standard low-pass, weighted-average filter, often used in image blurring, with the Gaussian standard deviation controlling the amount of smoothing. Effectively reduces high-frequency noise but potentially broadens pulse edges and overestimating rise-time \cite{gauss_filter}. The SG filter smooths data by fitting a small polynomial curve to a short window of neighboring points, then using that fitted curve to estimate the middle point. It moves this window along the data or, waveform and repeats the process, so the overall signal becomes smoother while keeping peaks and shapes better than a simple average filter~\cite{SG_2002,SG_2022}. The low-pass Butterworth filter acts like a smooth gate for signals. It lets the low frequencies pass through and gradually blocks the higher ones, with as little distortion as possible in the passband~\cite{Butterworth_1930}. It suppresses high-frequency noise with a maximally flat passband. Wavelet denoising filter, operates in the time–frequency domain~\cite{SG_wavelet_comparison}. Its successive separation of a signal into approximation and detail components using low-pass and high-pass filtering, enables multiscale analysis of local features, effectively preserving sharp features of the pulses.

\begin{figure}
    \centering
    \includegraphics[width=0.6\linewidth]{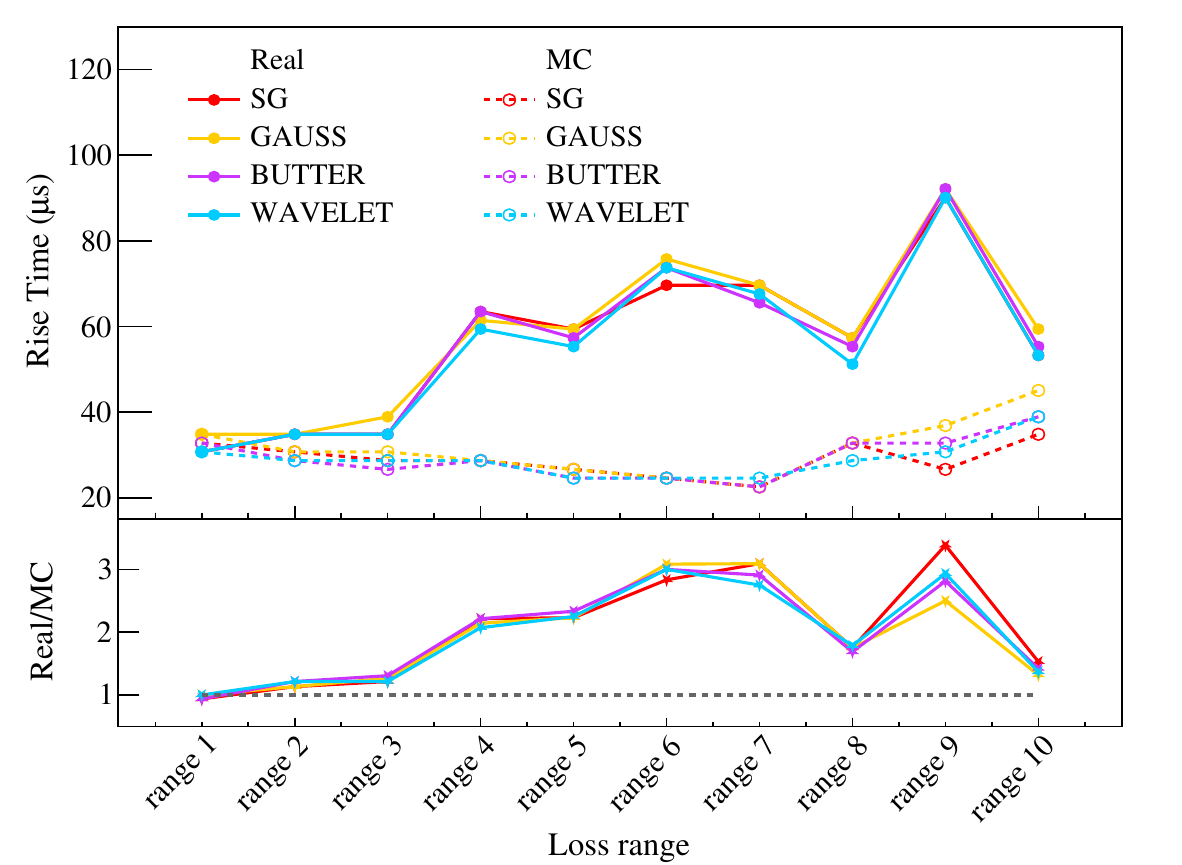}
    \caption{\textit{Upper panel}: Distribution of rise-time across the full range of CVAE loss values for both real and MC templates, shown for the four different filters. \textit{Bottom panel}: Ratio of rise-time in real data to MC as a function of CVAE loss over the same ranges.}
    \label{fig:rise_time_mc_real}
\end{figure}

Example filtered pulses for both real and MC data are shown in Fig.~\ref{fig:filtered_pulse_fake_real}. The complete datasets are processed with each filter, and the rise-time, that is, the time traversed to reach 90\% of the amplitude from 10\%, is evaluated after normalization. The rise-time as a function of CVAE loss score is presented in Fig.~\ref{fig:rise_time_mc_real} for different processing filters. For MC templates, the rise-time is typically confined to the range of 20–40~$\mu$s. In contrast, for real data, the rise-time increases with loss score, showing a significant enhancement beyond a loss score of $\sim$0.04. In the range 0.06–0.08, the rise-time is approximately three times larger than that of MC events. At higher loss values, the rise-time decreases with noticeable fluctuations, likely due to limited statistics in those regions.

\subsection{Discriminating LEE events \label{sec4.3:le_discrimination}}
\begin{figure}
    \centering
    \includegraphics[width=0.9\linewidth]{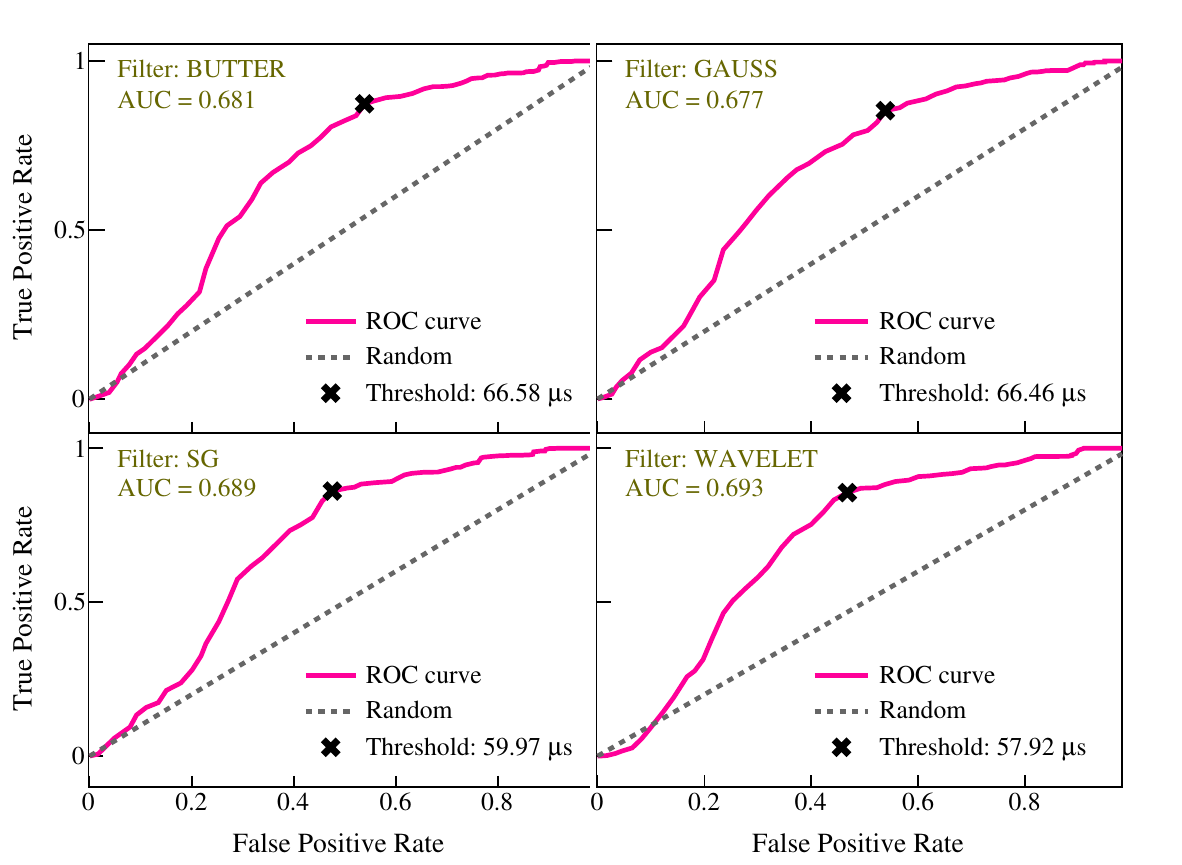}
    \caption{ROC curves used to determine the optimal rise-time threshold for discriminating LEE events. The evaluation is performed on a test dataset comprising 10,000 MC events and 633 real events with CVAE loss greater than 0.04. The area under the curve (AUC) is approximately 0.7 for all four filters. The dashed blue line represents a random classifier, while the black cross marks the optimal operating point corresponding to the selected rise-time threshold (TPR and FPR).}
    \label{fig:roc}
\end{figure}

Here, we estimate a rise-time-based threshold that can be applied during preprocessing to reduce the need to process all events through the computationally expensive ML pipeline in future analyses. For this purpose, we construct a mixed dataset comprising 10,000 MC events (labeled as signal-like) and 633 real events with CVAE loss $> 0.04$ (labeled as LEE events).
Template-level comparisons (Fig.~\ref{fig:anomaly_temp_real_fake}) show the emergence of a slow pre-pulse component that predominantly appears in this loss region. Since these events already pass standard data quality cuts, the high-loss population is unlikely to arise predominantly from random noise or misreconstructed events. Instead, it points to a distinct class of physical events contributing to the LEE.

To determine the optimal threshold, we use three standard metrics: recall, precision, and the $F_1$ score (harmonic mean), defined as
\begin{equation}
    \text{Recall} = \frac{\mathrm{TPR}}{\mathrm{TPR} + \mathrm{FNR}},
    \label{eq:recall}
\end{equation}
where TPR is the true positive rate, i.e. the rate of successfully predicted signals, and FNR is the false negative rate, i.e. signals that are predicted as LEE events.
\begin{equation}
    \text{Precision} = \frac{\mathrm{TPR}}{\mathrm{TPR} + \mathrm{FPR}},
    \label{eq:precision}
\end{equation}
where FPR is the false positive rate, i.e. LEE events predicted as signals.
\begin{equation}
    F_1 = \frac{2 \times (\text{Precision} \times \text{Recall})}{\text{Precision} + \text{Recall}}.
    \label{eq:f1}
\end{equation}

Recall is the ratio of correctly predicted signals to all the signals (MC events), and precision is the ratio of correctly predicted observations to the total number of predicted signals. However, instead of optimizing recall or precision individually, which can introduce bias, the $F_1$ score is used as it balances both metrics. The rise-time range of all the events is divided into 200 bins, and the $F_1$ score is computed for each threshold. Each threshold corresponds to a pair of TPR and FPR values, forming the Receiver Operating Characteristic (ROC) curve, as shown in Fig.~\ref{fig:roc} for all four filters.

The optimal operating point, corresponding to the maximum $F_1$ score, is indicated in Fig.~\ref{fig:roc}, along with the associated rise-time threshold. The area under the ROC curve (AUC) is found to be $\sim$0.7 for all filters, indicating moderate discrimination power using features of the waveform and CVAE. More studies are in progress to improve the discriminating power.

The optimized threshold values and corresponding performance metrics are summarized in Table~\ref{tab:filter_efficiency}.
\begin{figure}
    \centering
    \includegraphics[width=0.6\linewidth]{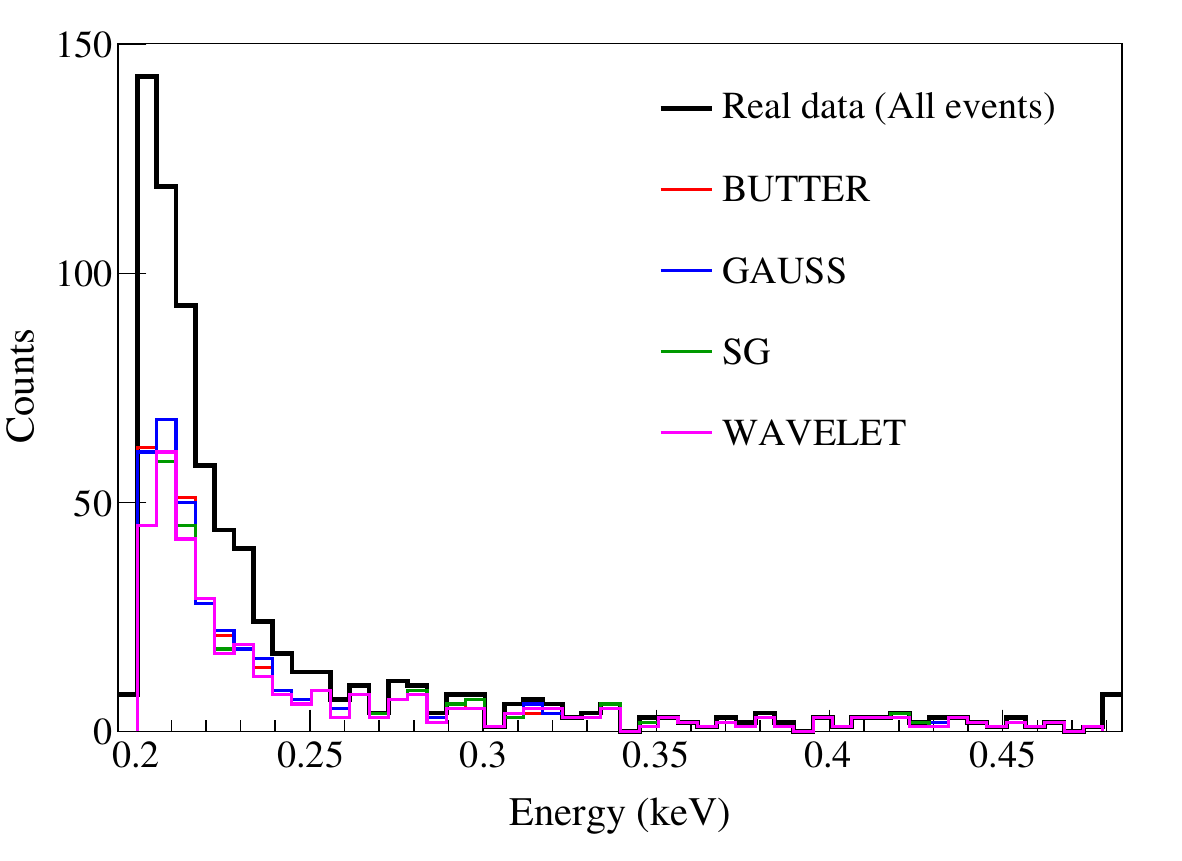}
    \caption{Comparison of energy spectra before and after applying rise-time–based selection thresholds for different filtering techniques, illustrating the impact of LEE suppression.}
    \label{fig:energy_spectra}
\end{figure}
\begin{table}[h!]
\centering
\small
\begin{tabular}{lcccccc}
\hline
\multirow{2}{*}{\textbf{Filter}} 
& \textbf{Threshold} & \multirow{2}{*}{$F_1$} & \multirow{2}{*}{AUC} 
& \textbf{TPR} & \textbf{FPR} & \textbf{LEE Rejection} \\
& ($\mu$s) & & & (\%) & (\%) & (\%) \\
\hline
SG      & 59.97 & 0.7365 & 0.689 & 86.06 & 47.55 & 52.45 \\
GAUSS   & 66.46 & 0.7134 & 0.677 & 85.31 & 53.87 & 46.13 \\
BUTTER  & 66.58 & 0.7239 & 0.681 & 87.34 & 53.87 & 46.13 \\
WAVELET & 57.92 & 0.7366 & 0.693 & 85.56 & 46.76 & 53.24\\
\hline
\end{tabular}
\caption{\label{tab:filter_efficiency}
Optimized rise-time thresholds and corresponding performance metrics for different filtering techniques.
}
\end{table}

Using the thresholds listed in Table~\ref{tab:filter_efficiency}, the selection criteria are applied to the real dataset. The resulting energy spectra before and after the selection are shown in Fig.~\ref{fig:energy_spectra}. Among all filters, the wavelet-based method performs best, achieving $\sim$53\% LEE rejection with an $F_1$ score of 0.74. This suggests that rise-time–based selections can help mitigate LEE contamination and may enhance the sensitivity of rare event search experiments near the detection threshold.

The robustness of the rise-time threshold is achieved by using these multiple filtering techniques (SG, Gaussian, Butterworth, and wavelet), and consistent trends are observed across all methods. The spread in the extracted rise-time values is incorporated as a systematic uncertainty when estimating the sensitivity in Sec.~\ref{sec4.4:sensitivity}. In addition, the stability of the CVAE training was verified by varying the train–validation split, yielding consistent results within statistical uncertainties.

\subsection{Optimizing sensitivity in CE\texorpdfstring{$\nu$}{nu}NS measurements\label{sec4.4:sensitivity}}
\begin{figure}
    \centering
    \includegraphics[width=\linewidth]{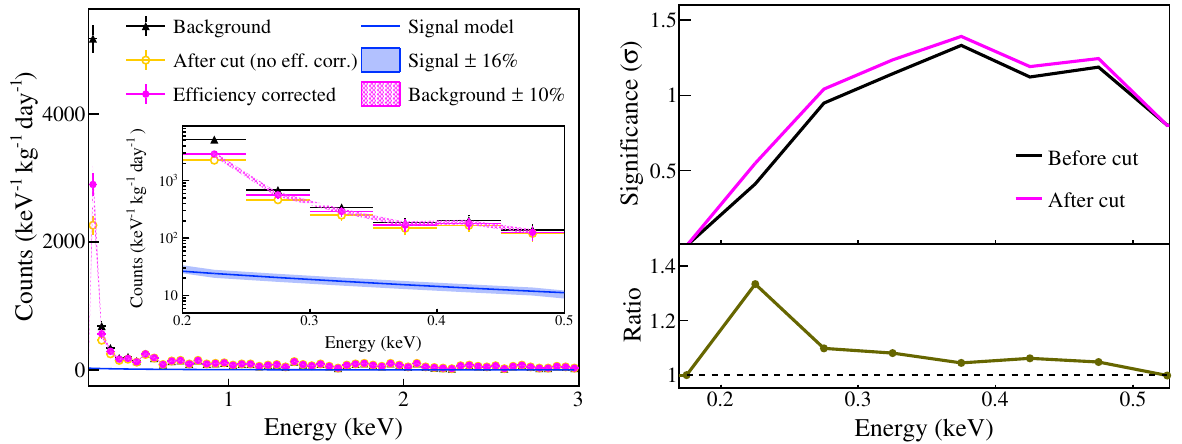}
    \caption{\textit{Left panel}: Differential rate unit (DRU, $\mathrm{keV^{-1} kg^{-1} day^{-1}}$) background spectra before and after applying the threshold cut using the optimized wavelet filtering algorithm, shown with and without efficiency correction. The signal model is overlaid in blue, along with its corresponding systematic uncertainty band. An additional 10\% systematic uncertainty is included for the background model. The inset highlights the LEE dominated region between 200–500 eV.
    \textit{Right panel}: Bin-wise significance in the 200–500 eV range before and after applying the rise-time cut. The bottom panel shows the corresponding ratio plot.}
    \label{fig:cenns_sensitivity}
\end{figure}

Finally, we estimate the improvement in significance for the CE$\nu$NS search in MINER using the formulated rise-time selection criteria to reject LEE events with slow components. We adopt a background spectrum from Sec.~\ref{sec4.3:le_discrimination} with an event rate of 400 $\mathrm{kg^{-1}day^{-1}}$ in the region of interest (ROI) of 200 eV to 3 keV. After applying the rise-time cut, the efficiency-corrected background spectrum is reduced to 332 $\mathrm{kg^{-1}day^{-1}}$.

For the raw background spectrum, we assume a systematic uncertainty of 7\%, following the previous study \cite{MINER_CEvNS}. An additional overall 8\% systematic uncertainty is associated with the filtering algorithm, resulting in a total systematic uncertainty of 10\% after LEE rejection. The efficiency-corrected background spectrum, along with its systematic uncertainty band, is shown in the left plot of Fig.~\ref{fig:cenns_sensitivity}.

We adopt a CE$\nu$NS signal model corresponding to the HFIR (High Flux Isotope Reactor) at Oak Ridge National Laboratory, which is being proposed as the new site for the MINER experiment as described in \cite{MINER_CEvNS}. At HFIR, an expected signal rate is approximately 10 events per day per kg in the ROI. The left plot of Fig.~\ref{fig:cenns_sensitivity} shows the signal model overlaid on the background spectrum, including a 16\% systematic uncertainty band as discussed in \cite{MINER_CEvNS}. The inset of the left plot of Fig.~\ref{fig:cenns_sensitivity} shows the same background and signal spectra, zoomed in the range 200-500 eV, which is mainly the LEE dominated region.

The upper right panel of Fig.~\ref{fig:cenns_sensitivity} presents the bin-wise significance considering 30 kg-days of exposure, calculated from
\begin{equation}
Z = \frac{s}{\sqrt{s+b}},
\end{equation}
where $s$ and $b$ denote the signal and background events, respectively.

The lower panel of the right-hand side plot of Fig.~\ref{fig:cenns_sensitivity} shows the ratio of significances before and after applying the rise-time cut. Overall, an improvement is observed in the lower energy bins, most notably in the lowest bin, where the significance increases by approximately a factor of ~1.4, due to the relatively higher signal contribution and LEE rate suppression in that region. In general, an improvement of about 10\% in overall significance is achieved, increasing from $(2.8\pm0.4~(\mathrm{sys}))\sigma$ before LEE rejection to $(3.1\pm0.5~\mathrm{(sys))}\sigma$ after applying the selection criteria.
This quoted improvement in sensitivity is based on a simplified estimate of the statistical significance, evaluated in the region of interest. While a full likelihood-based analysis is beyond the scope of the present work, this approach provides a reasonable first-order estimate of the impact of LEE suppression on the experimental sensitivity near threshold.
\section{Concluding remarks\label{sec5:conclusion}}

In this study, we present the evidence of LEE in MINER data using a sapphire detector. Under both reactor-ON and reactor-OFF conditions, a sharp rise is observed after each warming cycle, followed by a gradual decay over time. 
In this work, we introduce a novel approach to identify LEE events using a deep learning–based analysis pipeline. The method employs a neural network that incorporates an unsupervised CVAE, which extracts features from input pulses by applying convolutional windows across the waveform. The encoder part learns features and compresses them into a lower-dimensional latent space. The decoder part of the network reconstructs the input pulse from that latent space by minimizing the reconstruction loss defined in Eq.~\ref{eq:loss_tot}. Consequently, well-reconstructed pulses yield lower loss scores, while poorly reconstructed pulses yield higher loss scores.

We apply a pre-trained CVAE model, trained on MC (ideal) data, to a LEE-dominated dataset in the energy range of 200–500 eV. Our analysis reveals the presence of a slow component in the pre-pulse region for LEE events, which affects the rise-time of the pulses. Across different loss score intervals, the rise-times of MC pulses are predominantly concentrated between 20–40~$\mu$s. In contrast, for events with loss scores above 0.04, the rise-time of real data pulses increases to more than three times that of the MC pulses between 0.06--0.08 loss range.

This is the first observation of a deviation in pulse templates within the LEE-dominated region. Presuming that this rise-time difference serves as a discriminating feature for LEE events, we develop and demonstrate a pipeline to separate them from the dataset. We further compare different filter-based methods for estimating rise-time thresholds, finding that a wavelet-based filter provides the best performance, achieving approximately 53\% rejection of LEE events. For the upcoming MINER experiment at HFIR, the implementation of the proposed selection procedure enhances CE$\nu$NS sensitivity, delivering a 10\% increase in significance in the low-energy bin where the signal is most abundant. The approach developed here is directly relevant to other cryogenic experiments such as SuperCDMS, CRESST, EDELWEISS, and NUCLEUS, which report similar LEE rates. The proposed deep learning–based pulse-shape analysis provides a transferable and largely model-independent tool for identifying LEE backgrounds. Its adoption could enhance low-energy sensitivity and improve background understanding across these experiments.

Additionally, in the context of the hypotheses discussed in Sec.~\ref{sec1:intro}, our observations are consistent with scenarios involving bulk-related defects or microfractures formed during the cooling process \cite{microfracture_LEE,fracture_process}. At room temperature, the crystal may host intrinsic defects that exist in metastable configurations. As the detector is cooled, these defects gradually relax toward their ground state, releasing energy as low-energy phonons. This mechanism effectively leads to a delayed detector response arising from scattering and quasi-diffusive transport processes.
An alternative but related scenario involves the formation of microfractures during the initial stages of the cooling cycle due to thermal stress. These microfractures can act as transient energy-release sites, emitting phonons as they stabilize over time. Such processes would contribute to the time-dependent behavior of the LEE, including its decay after cooldown and regeneration following subsequent warm-up and recooling cycles.

In contrast, energy deposition from particle interactions is highly localized and produces athermal ballistic phonons, leading to a characteristically sharp rise in the pulse. The clear distinction in pulse shape between these two cases supports the interpretation that a significant fraction of the observed LEE events originates from bulk defects or microfracture-related processes.
However, it is important to emphasize that LEE is likely a multi-component phenomenon. While the observed pulse-shape deviations and time dependence are consistent with defect- or microfracture-related origins, additional contributions, such as surface-related effects or other detector-specific processes, cannot be excluded and may play a role, particularly at the lowest energies.

Although the application of rise-time based threshold reduces LEE events by up to 53\%, a pronounced excess persists in the lower energy bins, as shown in Fig.~\ref{fig:energy_spectra}. This residual excess suggests that additional contributions may arise from surface-related processes, for which we do not expect any slow components. Events originating near the sensor are expected to produce relatively faster rise-times due to reduced phonon propagation distances and limited scattering, and may therefore not be efficiently discriminated within the current analysis framework. A robust separation of bulk and surface contributions would require a dedicated analysis, which is beyond the scope of this study and is left for future work.

The methods and observations presented in this work are not specific to the MINER experiment and are expected to be broadly applicable to other low-threshold cryogenic detectors facing similar low-energy excess challenges. In particular, pulse-shape–based discrimination, combined with data-driven anomaly detection, offers a promising pathway toward mitigating detector-related backgrounds and enhancing sensitivity in next-generation rare-event searches.
\section{Acknowledgement}
This work was supported by the U.S. Department of Energy (DOE) under Grant Nos. DE-SC0018981 and DE-SC0017859. The authors gratefully acknowledge seed funding from the Mitchell Institute, which supported early conceptual development, prototyping, and operating costs for this experiment. We thank Texas A\&M's Nuclear Science Center management and staff for providing access to the reactor, facilities, and technical resources. Additional support was provided by the Department of Atomic Energy (DAE) and the Department of Science and Technology (DST), Government of India. This work is partially funded through the J.C. Bose Fellowship of the Anusandhan National Research Foundation (ANRF) awarded to B. Mohanty. The Australian Research Council supports J. L. Newstead's contributions through the ARC Centre of Excellence for Dark Matter Particle Physics (CE200100008). We also acknowledge the use of the Kanaad HPC cluster at SPS, NISER. 

\appendix

\bibliographystyle{JHEP}
\bibliography{biblio}

@article{MINER_CEvNS,
      title={CE$\nu$NS Search with Cryogenic Sapphire Detectors at MINER: Results from the TRIGA reactor data and Future Sensitivity at HFIR}, 
      author={Mondal, D. and others},
      year={2025},
      eprint={2510.15999},
      archivePrefix={arXiv},
      primaryClass={nucl-ex},
      url={https://arxiv.org/abs/2510.15999},
      journal={arXiv preprint},
}

@article{excess_workshop,
    author = "Adari, Prakruth and others",
    editor = "Fuss, A. and Kaznacheeva, M. and Reindl, F. and Wagner, F.",
    title = "{EXCESS workshop: Descriptions of rising low-energy spectra}",
    eprint = "2202.05097",
    archivePrefix = "arXiv",
    primaryClass = "astro-ph.IM",
    reportNumber = "FERMILAB-CONF-22-208-PPD-SCD-V",
    doi = "10.21468/SciPostPhysProc.9.001",
    journal = "SciPost Phys. Proc.",
    volume = "9",
    pages = "001",
    year = "2022"
}

@article{LEE_annual_review,
    author = "Baxter, Daniel and Essig, Rouven and Hochberg, Yonit and Kaznacheeva, Margarita and von Krosigk, Belina and Reindl, Florian and Romani, Roger K. and Wagner, Felix",
    title = "{Low-Energy Backgrounds in Solid-State Phonon and Charge Detectors}",
    eprint = "2503.08859",
    archivePrefix = "arXiv",
    primaryClass = "physics.ins-det",
    reportNumber = "FERMILAB-PUB-24-0953-ETD-PPD",
    doi = "10.1146/annurev-nucl-121423-100849",
    journal = "Ann. Rev. Nucl. Part. Sci.",
    volume = "75",
    number = "1",
    pages = "301--326",
    year = "2025"
}

@article{LEE_al_relaxation,
    author = "Romani, Roger K.",
    title = "{Aluminum relaxation as the source of excess low energy events in low threshold calorimeters}",
    eprint = "2406.15425",
    archivePrefix = "arXiv",
    primaryClass = "physics.ins-det",
    doi = "10.1063/5.0222654",
    journal = "J. Appl. Phys.",
    volume = "136",
    number = "12",
    pages = "124502",
    year = "2024"
}

@article{LEE_TESSERACT_bulk,
    author = "Chang, C. L. and others",
    collaboration = "TESSERACT",
    title = "{Spontaneous generation of athermal phonon bursts within bulk silicon causing excess noise, low energy background events, and quasiparticle poisoning in superconducting sensors}",
    eprint = "2505.16092",
    archivePrefix = "arXiv",
    primaryClass = "physics.ins-det",
    doi = "10.1063/5.0281876",
    journal = "Appl. Phys. Lett.",
    volume = "127",
    number = "26",
    pages = "263502",
    year = "2025"
}

@article{TESSERACT_2025,
    author = "Bui, T. K. and others",
    collaboration = "TESSERACT",
    title = "{First Limits on Light Dark Matter Interactions in a Low Threshold Two-Channel Athermal Phonon Detector from the TESSERACT Collaboration}",
    eprint = "2503.03683",
    archivePrefix = "arXiv",
    primaryClass = "hep-ex",
    doi = "10.1103/hsrl-crvf",
    journal = "Phys. Rev. Lett.",
    volume = "135",
    number = "16",
    pages = "161002",
    year = "2025"
}

@article{TESSERACT_2026,
    author = "Armatol, A. and others",
    collaboration = "TESSERACT",
    title = "{Low Energy Phonon Bursts Created By Fast Neutron Damage}",
    eprint = "2603.17964",
    archivePrefix = "arXiv",
    journal = "arXiv preprint",
    primaryClass = "physics.ins-det",
    month = "3",
    year = "2026"
}

@article{LEE_nucleus_2026,
    author = "Abele, H. and others",
    title = "{Characterization of the Low Energy Excess using a NUCLEUS $Al_2O_3$ detector}",
    eprint = "2603.07687",
    archivePrefix = "arXiv",
    journal = "arXiv preprint",
    primaryClass = "physics.ins-det",
    month = "3",
    year = "2026"
}

@article{LEE_CRESST_confpros,
	title={{Latest observations on the low energy excess in CRESST-III}},
	author={G. Angloher and others},
	journal={SciPost Phys. Proc.},
	pages={013},
	year={2023},
	publisher={SciPost},
	doi={10.21468/SciPostPhysProc.12.013},
	url={https://scipost.org/10.21468/SciPostPhysProc.12.013},
}

@article{CRESST_LEE_2024,
	author = {Angloher, G. and  others},
    collaboration = "CRESST",
	title = {DoubleTES detectors to investigate the CRESST low energy background: results from above-ground prototypes},
	DOI= "10.1140/epjc/s10052-024-13282-8",
	url= "https://doi.org/10.1140/epjc/s10052-024-13282-8",
	journal = {Eur. Phys. J. C},
	year = 2024,
	volume = 84,
	number = 10,
	pages = "1001",
}

@article{LEE_SuperCDMS,
    author = "Albakry, M. F. and others",
    collaboration = "SuperCDMS",
    title = "{Investigating the sources of low-energy events in a SuperCDMS-HVeV detector}",
    eprint = "2204.08038",
    archivePrefix = "arXiv",
    primaryClass = "hep-ex",
    reportNumber = "FERMILAB-PUB-22-496-PPD-SQMS",
    doi = "10.1103/PhysRevD.105.112006",
    journal = "Phys. Rev. D",
    volume = "105",
    number = "11",
    pages = "112006",
    year = "2022"
}

@article{LEE_CRESST_2019,
    author = "Abdelhameed, A. H. and others",
    collaboration = "CRESST",
    title = "{First results from the CRESST-III low-mass dark matter program}",
    eprint = "1904.00498",
    archivePrefix = "arXiv",
    primaryClass = "astro-ph.CO",
    doi = "10.1103/PhysRevD.100.102002",
    journal = "Phys. Rev. D",
    volume = "100",
    number = "10",
    pages = "102002",
    year = "2019"
}

@article{LEE_EDELWEISS_2019,
    author = "Armengaud, E. and others",
    collaboration = "EDELWEISS",
    title = "{Searching for low-mass dark matter particles with a massive Ge bolometer operated above-ground}",
    eprint = "1901.03588",
    archivePrefix = "arXiv",
    primaryClass = "astro-ph.GA",
    doi = "10.1103/PhysRevD.99.082003",
    journal = "Phys. Rev. D",
    volume = "99",
    number = "8",
    pages = "082003",
    year = "2019"
}

@article{EDELWEISS_2020,
    author = "Arnaud, Q. and others",
    collaboration = "EDELWEISS",
    title = "{First germanium-based constraints on sub-MeV Dark Matter with the EDELWEISS experiment}",
    eprint = "2003.01046",
    archivePrefix = "arXiv",
    primaryClass = "astro-ph.GA",
    doi = "10.1103/PhysRevLett.125.141301",
    journal = "Phys. Rev. Lett.",
    volume = "125",
    number = "14",
    pages = "141301",
    year = "2020"
}

@article{EDELWEISS_2023,
    author = "Augier, C. and others",
    collaboration = "EDELWEISS",
    title = "{Tagging and localization of ionizing events using NbSi transition edge phonon sensors for dark matter searches}",
    eprint = "2303.02067",
    archivePrefix = "arXiv",
    primaryClass = "physics.ins-det",
    doi = "10.1103/PhysRevD.108.022006",
    journal = "Phys. Rev. D",
    volume = "108",
    number = "2",
    pages = "022006",
    year = "2023"
}

@article{Ricochet_2023,
    author = "Augier, C. and others",
    collaboration = "Ricochet",
    title = "{First demonstration of 30 eVee ionization energy resolution with Ricochet germanium cryogenic bolometers}",
    eprint = "2306.00166",
    archivePrefix = "arXiv",
    primaryClass = "astro-ph.IM",
    doi = "10.1140/epjc/s10052-024-12433-1",
    journal = "Eur. Phys. J. C",
    volume = "84",
    number = "2",
    pages = "186",
    year = "2024"
}

@article{DAMIC_2023,
    author = "Aguilar-Arevalo, A. and others",
    collaboration = "SENSEI, DAMIC-M, DAMIC",
    title = "{Confirmation of the spectral excess in DAMIC at SNOLAB with skipper CCDs}",
    eprint = "2306.01717",
    archivePrefix = "arXiv",
    primaryClass = "astro-ph.CO",
    reportNumber = "FERMILAB-PUB-23-256-PPD",
    doi = "10.1103/PhysRevD.109.062007",
    journal = "Phys. Rev. D",
    volume = "109",
    number = "6",
    pages = "062007",
    year = "2024"
}

@article{DAMIC_2021,
    author = "Aguilar-Arevalo, A. and others",
    collaboration = "DAMIC",
    title = "{Characterization of the background spectrum in DAMIC at SNOLAB}",
    eprint = "2110.13133",
    archivePrefix = "arXiv",
    primaryClass = "hep-ex",
    reportNumber = "FERMILAB-PUB-21-498-AE-E-QIS",
    doi = "10.1103/PhysRevD.105.062003",
    journal = "Phys. Rev. D",
    volume = "105",
    number = "6",
    pages = "062003",
    year = "2022"
}

@article{SENSEI_2025,
    author = "Adari, Prakruth and others",
    collaboration = "SENSEI",
    title = "{First Direct-Detection Results on Sub-GeV Dark Matter Using the SENSEI Detector at SNOLAB}",
    eprint = "2312.13342",
    archivePrefix = "arXiv",
    primaryClass = "astro-ph.CO",
    reportNumber = "YITP-SB-2023-30, FERMILAB-PUB-23-0824-CSAID-PPD",
    doi = "10.1103/PhysRevLett.134.011804",
    journal = "Phys. Rev. Lett.",
    volume = "134",
    number = "1",
    pages = "011804",
    year = "2025"
}

@article{CONNIE_2024,
    author = "Aguilar-Arevalo, Alexis A. and others",
    collaboration = "CONNIE",
    title = "{Searches for CE{\ensuremath{\nu}}NS and Physics beyond the Standard Model using Skipper-CCDs at CONNIE}",
    eprint = "2403.15976",
    archivePrefix = "arXiv",
    journal = "arxiv preprint",
    primaryClass = "hep-ex",
    reportNumber = "FERMILAB-PUB-24-0714-PPD",
    month = "3",
    year = "2024"
}

@article{nature_2024,
author = {Anthony-Petersen and others},
year = {2024},
month = {07},
pages = {},
title = {A stress-induced source of phonon bursts and quasiparticle poisoning},
volume = {15},
journal = {Nature Communications},
doi = {10.1038/s41467-024-50173-8}
}

@article{microfracture_LEE,
  title = {Spontaneous damage annealing reactions as a possible source of low energy excess in semiconductor detectors},
  author = {Nordlund, Kai and Kong, Fanhao and Djurabekova, Flyura and Heikinheimo, Matti and Tuominen, Kimmo and Kuronen, Antti and Mirabolfathi, Nader},
  journal = {Phys. Rev. Mater.},
  volume = {9},
  issue = {11},
  pages = {113603},
  numpages = {21},
  year = {2025},
  month = {Nov},
  publisher = {American Physical Society},
  doi = {10.1103/p7s5-4qjw},
  url = {https://link.aps.org/doi/10.1103/p7s5-4qjw}
}

@article{radiation_LEE,
  title = {Sources of Low-Energy Events in Low-Threshold Dark-Matter and Neutrino Detectors},
  author = {Du, Peizhi and Egana-Ugrinovic, Daniel and Essig, Rouven and Sholapurkar, Mukul},
  journal = {Phys. Rev. X},
  volume = {12},
  issue = {1},
  pages = {011009},
  numpages = {45},
  year = {2022},
  month = {Jan},
  publisher = {American Physical Society},
  doi = {10.1103/PhysRevX.12.011009},
  url = {https://link.aps.org/doi/10.1103/PhysRevX.12.011009}
}

@article{CNN_CVAE_DARWIN,
    author = "Aalbers, J. and others",
    collaboration = "DARWIN",
    title = "{Model-independent searches of new physics in DARWIN with deep learning}",
    eprint = "2410.00755",
    archivePrefix = "arXiv",
    primaryClass = "physics.ins-det",
    doi = "10.1140/epjc/s10052-025-15161-2",
    journal = "Eur. Phys. J. C",
    volume = "86",
    number = "3",
    pages = "312",
    year = "2026"
}

@article{CNN_NEXT,
    author = "Kekic, M. and others",
    collaboration = "NEXT",
    title = "{Demonstration of background rejection using deep convolutional neural networks in the NEXT experiment}",
    eprint = "2009.10783",
    archivePrefix = "arXiv",
    primaryClass = "physics.ins-det",
    doi = "10.1007/JHEP01(2021)189",
    journal = "JHEP",
    volume = "01",
    pages = "189",
    year = "2021"
}

@article{CNN_XENON_2019,
    author = "Khosa, Charanjit K. and Mars, Lucy and Richards, Joel and Sanz, Veronica",
    title = "{Convolutional Neural Networks for Direct Detection of Dark Matter}",
    eprint = "1911.09210",
    archivePrefix = "arXiv",
    primaryClass = "hep-ph",
    doi = "10.1088/1361-6471/ab8e94",
    journal = "J. Phys. G",
    volume = "47",
    number = "9",
    pages = "095201",
    year = "2020"
}

@article{CNN_DM_halo_2019,
    author = "Bernardini, Mauro and Mayer, Lucio and Reed, Darren and Feldmann, Robert",
    title = "{Predicting dark matter halo formation in N-body simulations with deep regression networks}",
    eprint = "1912.04299",
    archivePrefix = "arXiv",
    primaryClass = "astro-ph.CO",
    doi = "10.1093/mnras/staa1911",
    journal = "Mon. Not. Roy. Astron. Soc.",
    volume = "496",
    number = "4",
    pages = "5116--5125",
    year = "2020"
}

@article{CVAE_LHC1_2021,
    author = "van Beekveld, Melissa and Caron, Sascha and Hendriks, Luc and Jackson, Paul and Leinweber, Adam and Otten, Sydney and Patrick, Riley and Ruiz De Austri, Roberto and Santoni, Marco and White, Martin",
    title = "{Combining outlier analysis algorithms to identify new physics at the LHC}",
    eprint = "2010.07940",
    archivePrefix = "arXiv",
    primaryClass = "hep-ph",
    doi = "10.1007/JHEP09(2021)024",
    journal = "JHEP",
    volume = "09",
    pages = "024",
    year = "2021"
}

@article{CVAE_LHC2_2021,
    author = "Cerri, Olmo and Nguyen, Thong Q. and Pierini, Maurizio and Spiropulu, Maria and Vlimant, Jean-Roch",
    title = "{Variational Autoencoders for New Physics Mining at the Large Hadron Collider}",
    eprint = "1811.10276",
    archivePrefix = "arXiv",
    primaryClass = "hep-ex",
    doi = "10.1007/JHEP05(2019)036",
    journal = "JHEP",
    volume = "05",
    pages = "036",
    year = "2019"
}

@article{COUNSplus,
    author = "Ackermann, N. and others",
    title = "{Direct observation of coherent elastic antineutrino-nucleus scattering}",
    eprint = "2501.05206",
    archivePrefix = "arXiv",
    primaryClass = "hep-ex",
    doi = "10.1038/s41586-025-09322-2",
    journal = "Nature",
    volume = "643",
    pages = "1229--1233",
    year = "2025"
}

@article{CONNIE,
    author = "Aguilar-Arevalo, Alexis and others",
    collaboration = "CONNIE",
    title = "{Search for coherent elastic neutrino-nucleus scattering at a nuclear reactor with CONNIE 2019 data}",
    eprint = "2110.13033",
    archivePrefix = "arXiv",
    primaryClass = "hep-ex",
    doi = "10.1007/JHEP05(2022)017",
    journal = "JHEP",
    volume = "05",
    pages = "017",
    year = "2022"
}

@article{NUGEN,
doi = {10.1088/1674-1137/adb9c8},
url = {https://doi.org/10.1088/1674-1137/adb9c8},
year = {2025},
month = {may},
publisher = {Chinese Physical Society and the Institute of High Energy Physics of the Chinese Academy of Sciences and the Institute of Modern Physics of the Chinese Academy of Sciences and IOP Publishing Ltd
				},
volume = {49},
number = {5},
pages = {053004},
author = {Belov, V. and others },
collaboration = {NuGeN Collaboration},
title = {New constraints on coherent elastic neutrino–nucleus scattering by the $\nu$GeN experiment*},
journal = {Chinese Physics C},

}

@article{NUCLEUS_2019,
    author = "Angloher, G. and others",
    collaboration = "NUCLEUS",
    title = "{Exploring $\hbox {CE}\nu \hbox {NS}$ with NUCLEUS at the Chooz nuclear power plant}",
    eprint = "1905.10258",
    archivePrefix = "arXiv",
    primaryClass = "physics.ins-det",
    doi = "10.1140/epjc/s10052-019-7454-4",
    journal = "Eur. Phys. J. C",
    volume = "79",
    number = "12",
    pages = "1018",
    year = "2019"
}

@article{NUCLEUS_2025,
    author = "Abele, H. and others",
    collaboration = "NUCLEUS",
    title = "{Commissioning of the NUCLEUS Experiment at the Technical University of Munich}",
    eprint = "2508.02488",
    archivePrefix = "arXiv",
    primaryClass = "hep-ex",
    doi = "10.1103/c95p-8kh2",
    journal = "Phys. Rev. D",
    volume = "112",
    number = "7",
    pages = "072013",
    year = "2025"
}

@article{TEXONOresult,
  title = {New Limits on the Coherent Neutrino-Nucleus Elastic Scattering Cross Section at the Kuo-Sheng Reactor-Neutrino Laboratory},
  author = "Karmakar, S. and others",
  collaboration = {TEXONO Collaboration},
  journal = {Phys. Rev. Lett.},
  volume = {134},
  issue = {12},
  pages = {121802},
  numpages = {6},
  year = {2025},
  month = {Mar},
  publisher = {American Physical Society},
  doi = {10.1103/PhysRevLett.134.121802},
  url = {https://link.aps.org/doi/10.1103/PhysRevLett.134.121802}
}

@article{Ricochet_comissioning,
    author = "Armatol, A. and others",
    collaboration = "Ricochet",
    title = "{Characterization of mini-CryoCube detectors from the Ricochet experiment commissioning at the Institut Laue-Langevin}",
    journal="arXiv",
    eprint = "2507.22751",
    archivePrefix = "arXiv",
    primaryClass = "astro-ph.IM",
    month = "7",
    year = "2025"
}

@misc{autoencoders_2021,
      title={Autoencoders}, 
      author={Dor Bank and Noam Koenigstein and Raja Giryes},
      year={2021},
      eprint={2003.05991},
      archivePrefix={arXiv},
      primaryClass={cs.LG},
      url={https://arxiv.org/abs/2003.05991}, 
}

@inproceedings{AE_anomaly_detection,
author = {Sakurada, Mayu and Yairi, Takehisa},
title = {Anomaly Detection Using Autoencoders with Nonlinear Dimensionality Reduction},
booktitle = {Anomaly Detection Using Autoencoders with Nonlinear Dimensionality Reduction},
year = {2014},
isbn = {9781450331593},
publisher = {Association for Computing Machinery},
address = {New York, NY, USA},
url = {https://doi.org/10.1145/2689746.2689747},
doi = {10.1145/2689746.2689747},
pages = {4–11},
numpages = {8},
location = {Gold Coast, Australia QLD, Australia},
series = {MLSDA'14}
}

@misc{VAE_2022,
      title={Auto-Encoding Variational Bayes}, 
      author={Diederik P Kingma and Max Welling},
      year={2022},
      eprint={1312.6114},
      archivePrefix={arXiv},
      primaryClass={stat.ML},
      url={https://arxiv.org/abs/1312.6114}, 
}

@inproceedings{KL_divergence,
  author       = {Irina Higgins others},
  title        = {beta-VAE: Learning Basic Visual Concepts with a Constrained Variational
                  Framework},
  booktitle    = {5th International Conference on Learning Representations, {ICLR} 2017,
                  Toulon, France, April 24-26, 2017, Conference Track Proceedings},
  publisher    = {OpenReview.net},
  year         = {2017},
  url          = {https://openreview.net/forum?id=Sy2fzU9gl},
  bibsource    = {dblp computer science bibliography, https://dblp.org}
}

@misc{CVAE,
      title={A guide to convolution arithmetic for deep learning}, 
      author={Vincent Dumoulin and Francesco Visin},
      year={2018},
      eprint={1603.07285},
      archivePrefix={arXiv},
      primaryClass={stat.ML},
      url={https://arxiv.org/abs/1603.07285}, 
}

@misc{tensorflowl_2016,
      title={TensorFlow: Large-Scale Machine Learning on Heterogeneous Distributed Systems}, 
      author={Martín Abadi and others},
      year={2016},
      eprint={1603.04467},
      archivePrefix={arXiv},
      primaryClass={cs.DC},
      url={https://arxiv.org/abs/1603.04467}, 
}

@inproceedings{relu_2013,
  title={Rectifier Nonlinearities Improve Neural Network Acoustic Models},
  author={Maas, Andrew L. and Hannun, Awni Y. and Ng, Andrew Y.},
  booktitle={Proc. ICML Workshop on Deep Learning for Audio, Speech and Language Processing},
  year={2013}
}

@article{gauss_filter,
  title={Scale-Space for Discrete Signals},
  author={Lindeberg, Tony P.},
  journal={IEEE Transactions on Pattern Analysis and Machine Intelligence},
  volume={12},
  number={3},
  pages={234--254},
  year={1990}
}

@article{SG_2002,
    author = "Savitzky, Abraham. and Golay, M. J. E.",
    title = "{Smoothing and Differentiation of Data by Simplified Least Squares Procedures.}",
    doi = "10.1021/ac60214a047",
    journal = "Anal. Chem.",
    volume = "36",
    number = "8",
    pages = "1627--1639",
    year = "2002"
}

@article{SG_2022,
  title={Why and How Savitzky–Golay Filters Should Be Replaced},
  author={Michael Schmid and David Rath and Ulrike Diebold},
  journal={ACS Measurement Science Au},
  year={2022},
  volume={2},
  pages={185 - 196},
  url={https://api.semanticscholar.org/CorpusID:246988058}
}

@article{Butterworth_1930,
  title = {On the {{Theory}} of {{Filter Amplifiers}}},
  volume = {7},
  journal = {Experimental Wireless \& the Wireless Engineer},
  author = {Butterworth, S.},
  month = {Oct},
  year = {1930},
  pages = {536-541}
}

@article{SG_wavelet_comparison,
title = {Wavelet denoising of Gaussian peaks: A comparative study},
journal = {Chemometrics and Intelligent Laboratory Systems},
volume = {34},
number = {2},
pages = {187-202},
year = {1996},
issn = {0169-7439},
doi = {https://doi.org/10.1016/0169-7439(96)00026-3},
url = {https://www.sciencedirect.com/science/article/pii/0169743996000263},
author = {C.R. Mittermayr and S.G. Nikolov and H. Hutter and M. Grasserbauer},
}

@article{fracture_process,
    author = {{\r{A}}str{\"o}m, J. and others},
    title = "{Fracture Processes Observed with A Cryogenic Detector}",
    eprint = "physics/0504151",
    archivePrefix = "arXiv",
    doi = "10.1016/j.physleta.2006.03.059",
    journal = "Phys. Lett. A",
    volume = "356",
    pages = "262--266",
    year = "2006"
}
\end{document}